\documentclass[english,aps,preprint]{revtex4}
\usepackage[T1]{fontenc}
\usepackage[latin1]{inputenc}
\usepackage{graphicx}
\usepackage{amssymb}
\usepackage{setspace}

\makeatletter

\usepackage{babel}
\makeatother
\begin{document}

\title{Adiabatic Charge Control in a Single Donor Atom Transistor}

\author{Enrico Prati}
\affiliation{Laboratorio MDM, IMM-CNR, Via Olivetti 2, I-20041 Agrate Brianza, Italy}
\email{enrico.prati@cnr.it}

\author{Matteo Belli}
\affiliation{Laboratorio MDM, IMM-CNR, Via Olivetti 2, I-20041 Agrate Brianza, Italy}
\author{Simone Cocco}
\affiliation{Laboratorio MDM, IMM-CNR, Via Olivetti 2, I-20041 Agrate Brianza, Italy}
\author{Guido Petretto}
\affiliation{Laboratorio MDM, IMM-CNR, Via Olivetti 2, I-20041 Agrate Brianza, Italy}
\author{Marco Fanciulli}
\affiliation{Laboratorio MDM, IMM-CNR, Via Olivetti 2, I-20041 Agrate Brianza, Italy}
\affiliation{Dipartimento di Scienza dei Materiali, Universit\`a degli Studi Milano-Bicocca, Via Cozzi 53, I-20125 Milano, Italy}

\begin{abstract}
We charge an individual donor with electrons stored in a quantum dot in its proximity.
A Silicon quantum device containing a single Arsenic donor and an electrostatic quantum dot in parallel is realized in a nanometric field effect transistor. The different coupling capacitances of the donor and the quantum dot with the control and the back gates are exploited to generate a relative rigid shift of their energy spectrum as a function of the back gate voltage, causing the crossing of the energy levels. We observe the sequential tunneling through the $D^{2-}$ and the $D^{3-}$ energy levels of the donor hybridized at the oxide interface at 4.2 K. Their respective states form an honeycomb pattern with the quantum dot states. It is therefore possible to control the exchange coupling of an electron of the quantum dot with the electrons bound to the donor, thus realizing a physical qubit for quantum information processing applications. 
\end{abstract}
\maketitle

\renewcommand{\baselinestretch}{2}
\section{Introduction}
We charge a single dopant lying in the channel of a Silicon nanometric field effect transistor (NanoFET) with electrons of the electrostatic quantum dot formed in the channel, in view of realizing tunable exchange coupling devices for atomic scale electronics and quantum information processing.

A groundbreaking step to improve the computing effectiveness and the logic algorithms beyond the CMOS technology consists of the transition from classical devices to atomic-scale quantum devices.\cite{Loss98,DiVincenzo00} In such devices the architecture can be constructed around suitable isolated dopant atoms,\cite{Sanquer09} capable to behave as quantum dots by switching from zero to few electrons occupation according to the atomic species and the applied electric field. An isolated atom, or few atoms deterministically positioned in a quantum device, may be used to store and to process information, and to switch between different charge, electron spin, and nuclear spin states. Donors in Silicon can realize those functionalities. \cite{Sanquer09,Rogge08} Single atom spintronics and solid state qubits are among the most relevant exploitation opportunities for the ultimate Si-based nanostructures in which a single donor is an essential part of the working principle of the device.  
Coulomb blockade of single donors in NanoFETs has been previously reported\cite{Rogge08,Prati08,Rogge10,Sanquer09,Tan10}
by studying the sequential tunneling through $D^0$ and $D^{-}$ states, also under microwave irradiation.\cite{Prati08}

By operating the sample in Coulomb blockade regime, we show that the electrons bound to a donor close to the oxide interface and those in another electrostatic quantum dot confined at the oxide interface are differently coupled respectively to the control gate and the back gate. We consequently tune the system among different charge states. In particular, a single electron is shifted from the quantum dot to the donor, where exchange coupling \cite{Friesen03} with other electrons bound to the donor becomes effective.
Such condition is achieved in the honeycomb region between the $D^{2-}$ and the $D^{3-}$ states of the donor at the interface \cite{Rogge08}, for which a mutual capacitive coupling with the quantum dot is observed, thus experimentally realizing a donor based physical qubit.
We conclude that the energy levels of a single donor close to an interface in a solid state quantum device can be controlled with respect to the quantum dot formed in the channel by means of two asymmetric parallel gates. 
We realize particle exchange between the donor with the quantum dot lying in its proximity.
The exchange coupling between the the electrons of the quantum dot and the electronic system of the donor can therefore be controlled, according to previously proposed qubit schemes \cite{Friesen03}.
\section{Experimental characterization}
\subsection{Samples and experimental setup}
\begin{figure}[b]
\begin{center}
\includegraphics[width=0.48\textwidth]{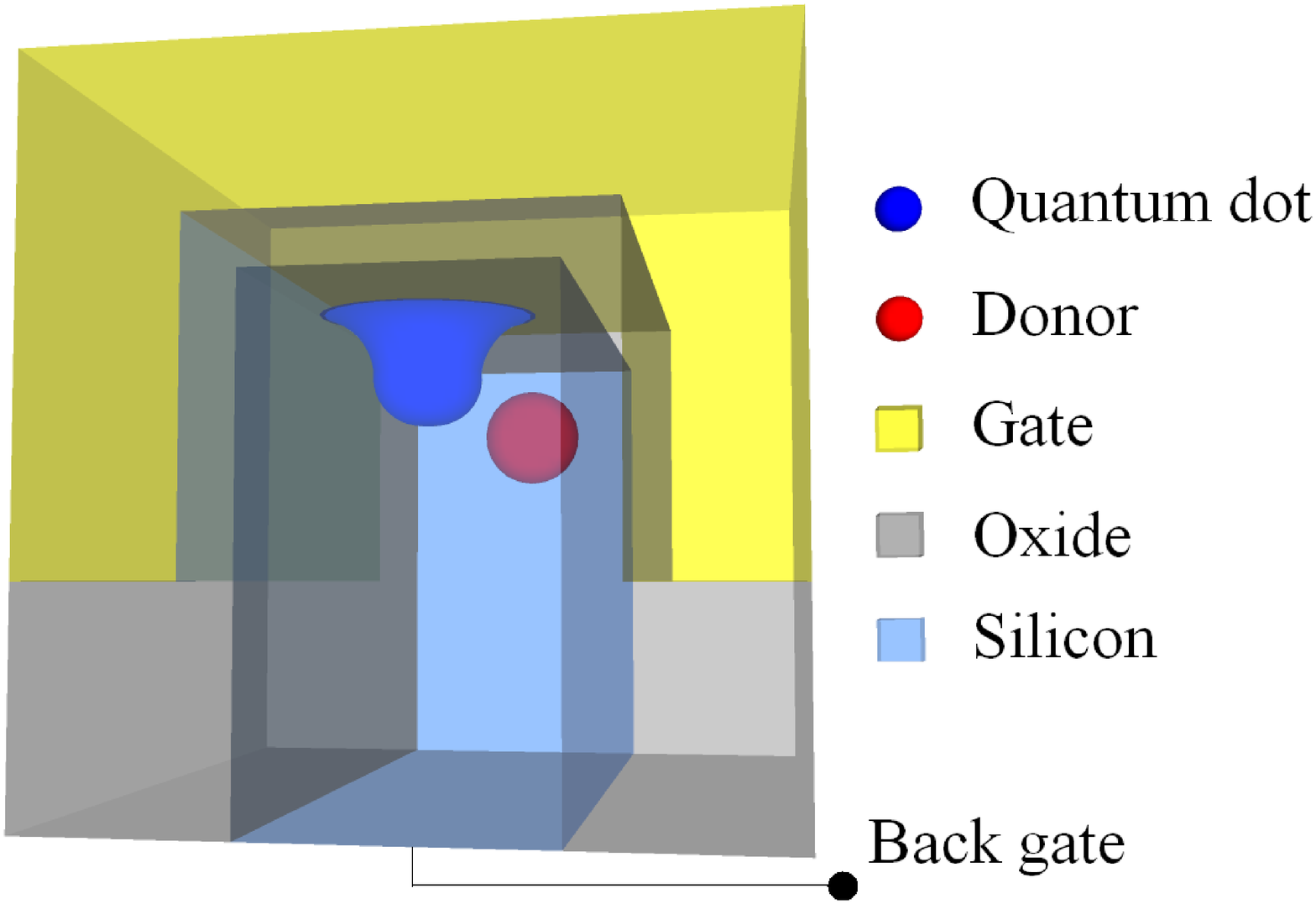}
\includegraphics[width=0.30\textwidth]{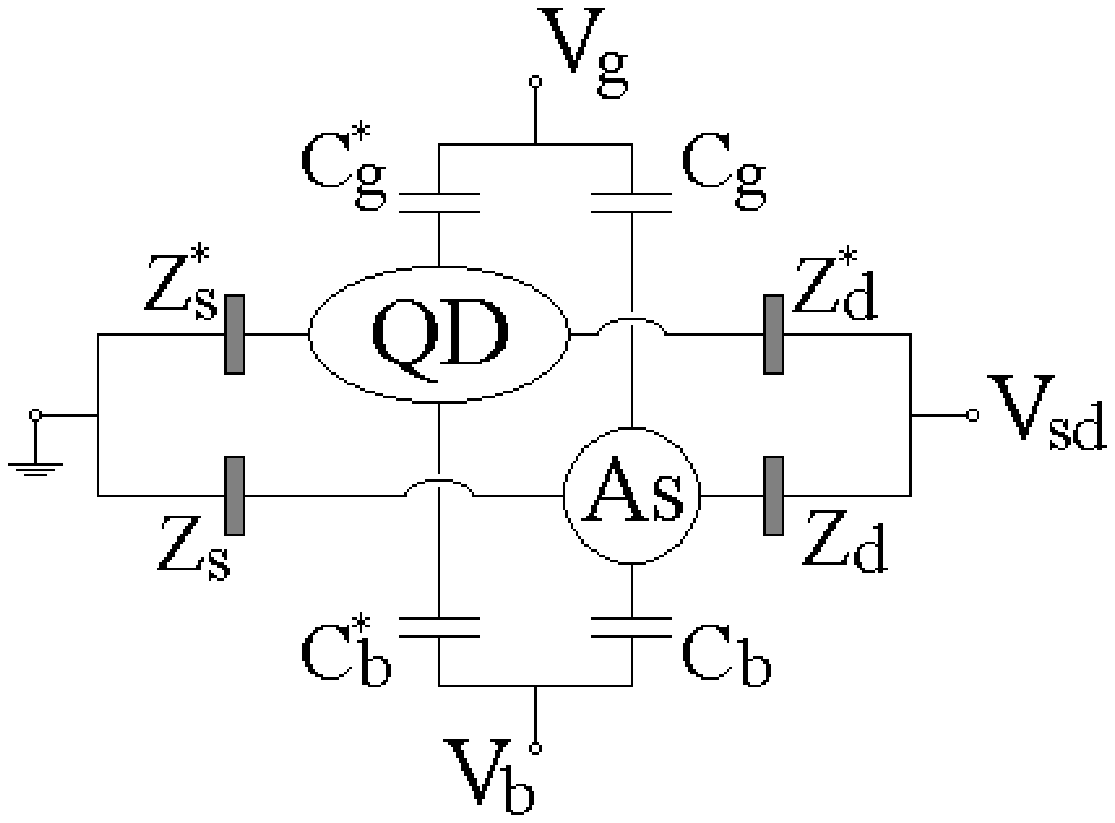}
\end{center}
\caption{\textbf{a.} A sketch of the n-type NanoFET device studied in this work when quantum effects become visible at 4 K. The width and length of the device channel are respectively 116 nm and 50 nm. The  gate is separated fom the Si channel by a 9.3 nm oxide layer, while ohmic contacts are ensured by highly As doped regions at the leads. \textbf{b.} Circuital scheme of the quantum dot and the donor capacitively coupled with the electrodes of the NanoFET. The source and the drain contacts are connected to the Arsenic donor quantum dot and to the quantum dot via an impedance $Z_{s,d}$ and $Z_{s,d}^{*}$ respectively which include the capacitances $C_{s,d}$ and $C_{s,d}^{*}$.}
\label{Figure1}
\end{figure}

The experiment was carried out in a n-channel Silicon field effect device having
a nominal length of $116$ nm and a width of $50$ nm, with gate oxide thickness of
$9.3$ nm. The source and the drain contacts are obtained by As doping of Silicon. A sketch of the sample and the circuital scheme are shown in Figure \ref{Figure1}.
The presence of the donor in the quantum device is due to the diffusion of few donors from the source and the drain to the nanometric channel.\cite{Prati08,Rogge10} Only some samples exhibit the conduction due to one or few donors with energy levels below the conduction band edge. Several samples with similar features have been characterized. For the sake of clarity, here all the results refer to the same sample, in which a single donor has been identified.
The drain current $I_{\mathrm{ds}}$ was measured by a transimpedance amplifier with bandwidth from dc to about $10$ kHz powered by independent batteries, while the sample was controlled by a PXI system providing output voltages with a resolution better than $350$ $\mu$V. We measured the sample immersed in liquid He at a temperature of $4.2$ K. 
%
\subsection{Coulomb blockade characterization}
From the current-voltage characteristics $I_{\mathrm{ds}}(V_{g})$ of the device between $4.2$ K and $90$ K we measured a threshold gate voltage $V_{\mathrm{T}}=2.530$ V at $V_{b} =$\linebreak $0$ V at $4.2$ K . Below about $40$ K the isolated conductance peaks are visible, but it is generally possible to determine the threshold of the background current by extrapolating the limiting value from the temperature trend.  
Modified equations with respect of the standard treatment of the Coulomb blockade problem \cite{Beenakker91,Sze3rded} have been used in order to include the back gate capacitance C$_{\mathrm{b}}$. The positive and negative slopes of the Coulomb blockade diamond can thus be expressed as $\frac{C_{\mathrm{G}}}{C_{\mathrm{\Sigma}}-C_{\mathrm{d}}}$ and $-\frac{C_G}{C_{d}}$, where $C_\Sigma = C_{s} + C_{d} + C_{g
} + C_{b}$ and where
$C_{s}$, $C_{d}$ and $C_{g}$ are the coupling capacitances between either the quantum dot or the donor and source, drain and gate contacts respectively (Figure 1b). 
The charging energy maintains the expression $E_{ch} = \frac{\mathrm{q}}{C_\Sigma}$ like in the single gate case. 

\begin{figure}[t]
\begin{center}
\includegraphics[width=0.48\textwidth]{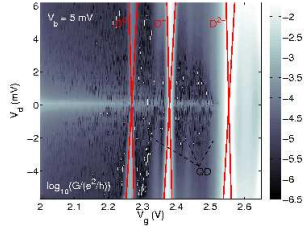}
\end{center}
\caption{Stability diagram of the logarithmic conductance $log_{10}(G/\frac{e^2}{h})$ at $4.2$ K and $V_{\mathrm{b}}=5$ mV. The band around $V_{\mathrm{d}} = 0$ V is an artifact due to the calculation of the conductance from the measured current value. 
} 
\label{Figure2}
\end{figure}

In Figure \ref{Figure2} the stability diagram in a voltage window including the threshold voltage is shown at a back gate voltage $V_{b}=5$ mV.
The first and third peaks correspond to the first ($N= 0\rightarrow 1 \rightarrow 0$) and the second ($N= 1 \rightarrow 2 \rightarrow 1$) charge state of the donor ($D^0$ and $D^{-}$ respectively). Several regularly spaced peaks emerge from the noise background in proximity of the conduction due to the donor and starting from around the threshold voltage.
The conversion from the applied control gate voltages to the corresponding energy shifts is obtained from the lever-arm factor $\alpha = 0.245$ in the case of the As donor and $\alpha = 0.220$ for the electrostatic quantum dot. The charging energy of the donor $U$ is 27.3 meV, while the spacing between the first peak and the conduction band edge $E_ {C}-E_1$ is about 64 meV. Such results are close to those previously reported,\cite{Rogge06,Prati08,Sanquer09} as expected for the As donors used in the source and drain contacts. 

\subsection{Single donor and quantum dot spectrum dependence on the back gate}
\begin{figure}[t]
\begin{center}
\includegraphics[width=0.45\textwidth]{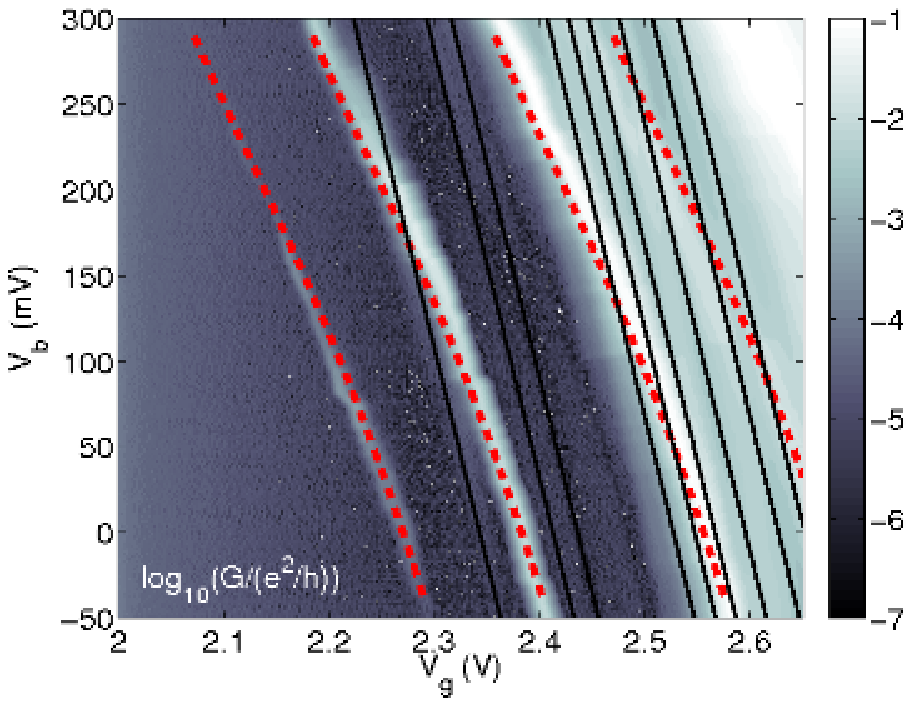}
\includegraphics[width=0.55\textwidth]{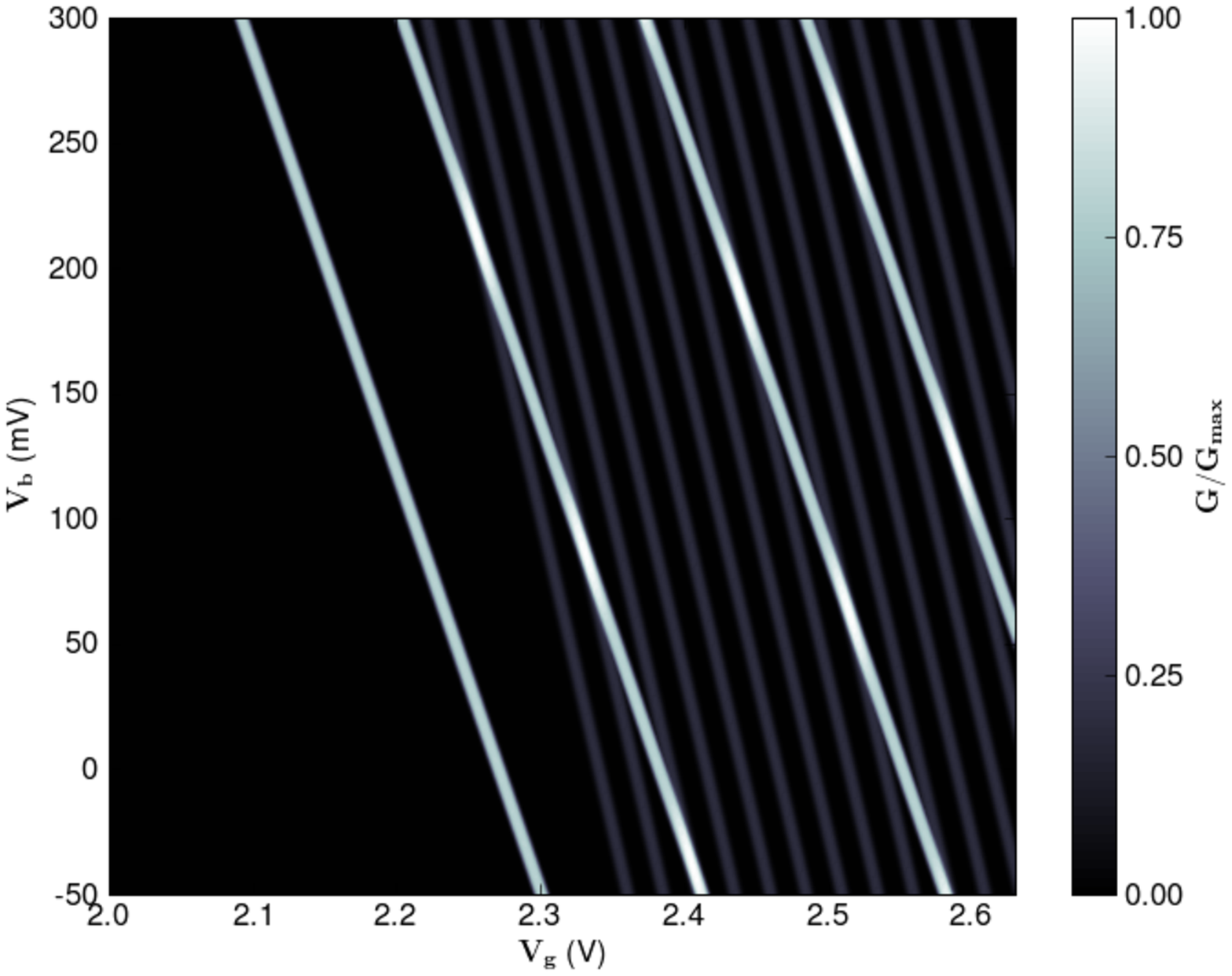}
\end{center}
\caption{\textbf{a.} Stability diagram of $log_{10}(G/\frac{e^2}{h})$ as a function of both the control gate and the back gate potentials at constant drain-source bias $V_{\mathrm{ds}}$ at $4.2$ K. The black lines identify the conductance peaks due to the transport through quantum dot levels, while the red curves are a guide for the eye to recognize donor conductance peaks at different back gate voltage values. The quantum dot levels are roughly evenly spaced, as expected, though they are relatively weak. They can be clearly observed only either in proximity of donor conductance peaks or above the threshold $V_T$ of the device. At low back gate potential the conductance values are identical to the corresponding $V_{d}$ slice of Figure \ref{Figure2}.   \textbf{b.} Simulation of the same system made of a donor and a quantum dot at $T=2$K. The simulation relies on the constant interaction model for the calculation of the total energy of the system, and the probabilities of occupation and the currents are obtained as stationary solution of the transition rate equation.
The capacitances were extracted from experimental data (Table I).
The conductance G of the two series of peaks refers to an arbitrary scale.}
\label{Figure3}
\end{figure}

\begin{figure}[t]
\begin{center}
\includegraphics[width=0.48\textwidth]{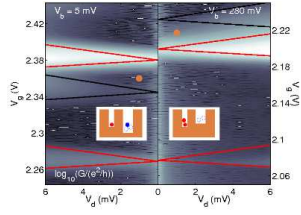}
\end{center}
\caption{Two different charge states of the double dot system, obtained by adjusting the voltages $V_{g}$ and $V_{b}$. The Coulomb blockade of the donor is represented with the red borders, the quantum dot in black.}
\label{Figure4}
\end{figure}

To attribute the Coulomb blockade peaks to the two quantum dots, we exploit the back gate control to which they are differently electrostatically coupled.
The peaks shown in the stability diagram of Figure \ref{Figure2} exhibit very similar slopes. In absence of further information it would not be possible to assign each of them either to a single quantum dot or to two separate quantum systems. In order to distinguish the peaks associated to the donor and to the quantum dot, we varied the back gate potential (Figure \ref{Figure3}). Two different interleaved series of peaks are discriminated as an effect of the different coupling to the back gate of two distinct electron systems lying in the channel. In Figure \ref{Figure3} a first set of roughly evenly spaced conductance peaks shifts linearly as a function of the $V_{b}$, while a second set of peaks is characterized by a different slope whose absolute value decreases at high back gate voltages. The former, because of the regular spacing, can be ascribed to a quantum dot due to electrostatic confinement of the electron wavefunction. The slope of the lines depicted in Figure \ref{Figure3} are, as a first order approximation, equal to $-\frac{C_{g}}{C_{b}}$.
The quantum dot is weakly coupled, as observed by the low conductance, so the peaks are detectable only when they occur at gate voltages either close to donor peaks or above the threshold. The set of peaks with larger spacing, by virtue of the correspondence of its charging energy and its binding energy with respect to the conduction band edge, are assigned to the energy levels of an isolated As donor, like in some other identical samples of the same batch.
The interpretation of the two gate stability diagram (Figure \ref{Figure3} \textbf{a}) in terms of the sequential tunneling through a quantum dot and a single donor has been confirmed by implementing an analytical model including both a metallic quantum dot and a donor quantum dot (Figure \ref{Figure3} \textbf{b}). Figure \ref{Figure3} \textbf{b} shows the simulation of a system made by a metallic quantum dot and a donor in parallel. 
The spectrum of the donor has been modeled as a spin (2-fold) degenerate ground state and a spin (2-fold) degenerate excited level separated from the former by the valley splitting of the interface case \cite{Rogge08} $\Delta_E$ = 15.4 meV extracted from the Figure \ref{Figure3} \textbf{a}, which has a similar origin of the valley-orbit splitting of the bulk case.\cite{Mayur93}
The coupling capacitances and the corresponding relevant parameters describing both donor and quantum dot are reported in Table \ref{Table1}.
The bending of the donor lines observed in Figure 3 \textbf{a} is associated to the effect of the back gate, which sligthly pulls the electron wavefunction away from the Si/SiO$_2$ interface when the voltage $V_{b}$ is increased. The consequent decrease of the gate capacitance caused by the larger distance is reflected by a smaller slope $-\frac{C_{g}}{C_{b}}$.
In order to better understand the experimental results, we explored capacitance couplings between the quantum dot and the donor. The lack of a triple point pair at the intersections of the two sets of peaks returns a negligible mutual capacitance at low gate voltages.
From Figure \ref{Figure3} \textbf{a} one observes that for suitable pairs $(V_{g},V_{b})$ the energy of the two electron systems are degenerate within the width of the peaks. At such crossing points the relative filling of the two quantum dots changes from $(N,M) \rightarrow (N,M+1) \rightarrow (N+1,M+1)$ as a function of the gate voltage to $(N,M) \rightarrow (N+1,M) \rightarrow (N+1,M+1)$ where $N$ and $M$ are the electron occupancy of the donor and the quantum dot respectively. The different filling sequence is shown in Figure \ref{Figure4} at two different back gate voltages starting from $N=1$. 
Operating the sample in the Coulomb blockade region at $N=1$ it is therefore possible to change the charge of the quantum dot and of the donor by just changing the back gate voltage.
\subsection{Adiabatic charge control}
We now turn to the study of the high front gate voltage regime, where it is possible to adiabatically shift an electron spin from the quantum dot site to the donor site where it couples with other electrons already bound to the donor. Indeed, when the number of electrons increases in the donor, the formation of a honeycomb pattern is observed, as shown in Figure 5 \textbf{a}. Such anticrossing is due to the mutual capacitance between the donor system and the quantum dot, while the tunnel coupling $t_c$ is negligible. The anticrossing corresponds to an energy gap which induces a voltage shift 
$ \Delta V_{{g(As)}} =\frac{C_{{As-QD}}}{C_{{As-QD}}C_{{g-QD}}+C_{{g-As}}C_{\Sigma{As}}} $
where $\Delta V_{{g(As)}}$ is the shift along $V_{g}$ of the Coulomb blockade pattern of the As donor, and the other capacitances are those connecting the quantum dot, the As donor and the front gate respectively, while $C_{\Sigma{As}}$ is the donor self capacitance. The anticrossing determines the condition to adiabatically shift the voltage in the $V_{g}-V_{b}$ plane to move a localized electron bound to the As donor, to the quantum dot site and \textit{viceversa}. In the proposed charge configuration, it is possible to adjust the gate voltages in order to tune the exchange coupling $J$ of an electron spin $\textbf{s}_{QD}$ of the quantum dot with the total spin  $\textbf{S}_{As}=\sum_{1}^{3} \textbf{s}_{As,i}$ of the three electrons lying in the donor site from a negligible ($J\approx0$) to the maximum value. The latter is obtained by driving the gate voltages from the region between the dashed lines on the right side of Figure 5 \textbf{a} to that at the left side. Figure 5 \textbf{b} shows the simulation of the honeycomb region obtained by adding a mutual capacitance of $C_{{As-QD}}=5$ aF.
Consequently, the device acts as a tuneable controller of the exchange coupling of individual spins, by shifting the position of a single electron from a site to another, one of whom is an individual donor.
\section{Conclusions}
To conclude, a quantum device constituted by a single As donor quantum dot and an electrostatic quantum dot in parallel, controlled by two asymmetric front and back gates, is exploited. The distinction between the energy levels of the As donor and the quantum dot is made possible by their different coupling to the back gate, which allows to identify two families of conductance peaks.
In addition to the $D^0$ and the $D^-$ states, the $D^{2-}$ and $D^{3-}$ charge state of the donor state hybridized at the oxide interface are observed, by virtue of the Coulomb blockade regime which, at sufficiently low temperature, eliminates the background conduction above the conduction band edge. The honeycomb pattern of the donor and the quantum dot states is observed at high front gate voltage. In that region we realize a physical qubit by adiabatically tuning the exchange coupling of an electron spin lying in the quantum dot with the total spin of the electrons bound to the donor quantum dot from zero to maximum coupling.
\begin{table}[h]
\caption{Coupling capacitances which describe the system according to the model depicted in Figure \ref{Figure1} at low front gate voltage. The QD-donor coupling capacitance at low front gate voltage is assumed negligible. The corresponding charging energy $E_{ch}$ and the $\alpha$ lever-arm parameter are also included, together with the valley splitting between the ground and the first excited state of the donor $\Delta E$.}
\label{Table1}
\begin{tabular}{lcc}
\hline
 & As donor & QD \\
\hline
$C_{s}$ (aF) & $1.74$ & $11.44$ \\
$C_{d}$ (aF) & $1.83$ & $8.76$ \\
$C_{g}$ (aF) & $1.44$ & $6.44$ \\
$C_{b}$ (aF) & $0.86$ & $2.57$ \\
$E_{ch}$ (meV) & $27.3$ & $5.5$ \\
$\alpha$ (meV/mV) & $0.245$ & $0.220$ \\
$\Delta E$ (meV) & $15.4$ & - \\
\hline
\end{tabular}
\end{table}



\begin{acknowledgments}
The authors would like to acknowledge P. Cappelletti (Numonyx) and STMicroelectronics for providing the samples; M. Sanquer and M. Pierre (CEA-Grenoble) for providing the software used to simulate the Coulomb blockade; A. Verduijn and S. Rogge for useful discussion of the results. This work has been partially supported by the European Unions seventh Framework (FP7 2007/2013) under the grant agreement nr: 214989-AFSiD project.
\end{acknowledgments}




\bibliography{achemso-demo}

\begin{figure}[t]
\begin{center}
\includegraphics[width=0.46\textwidth]{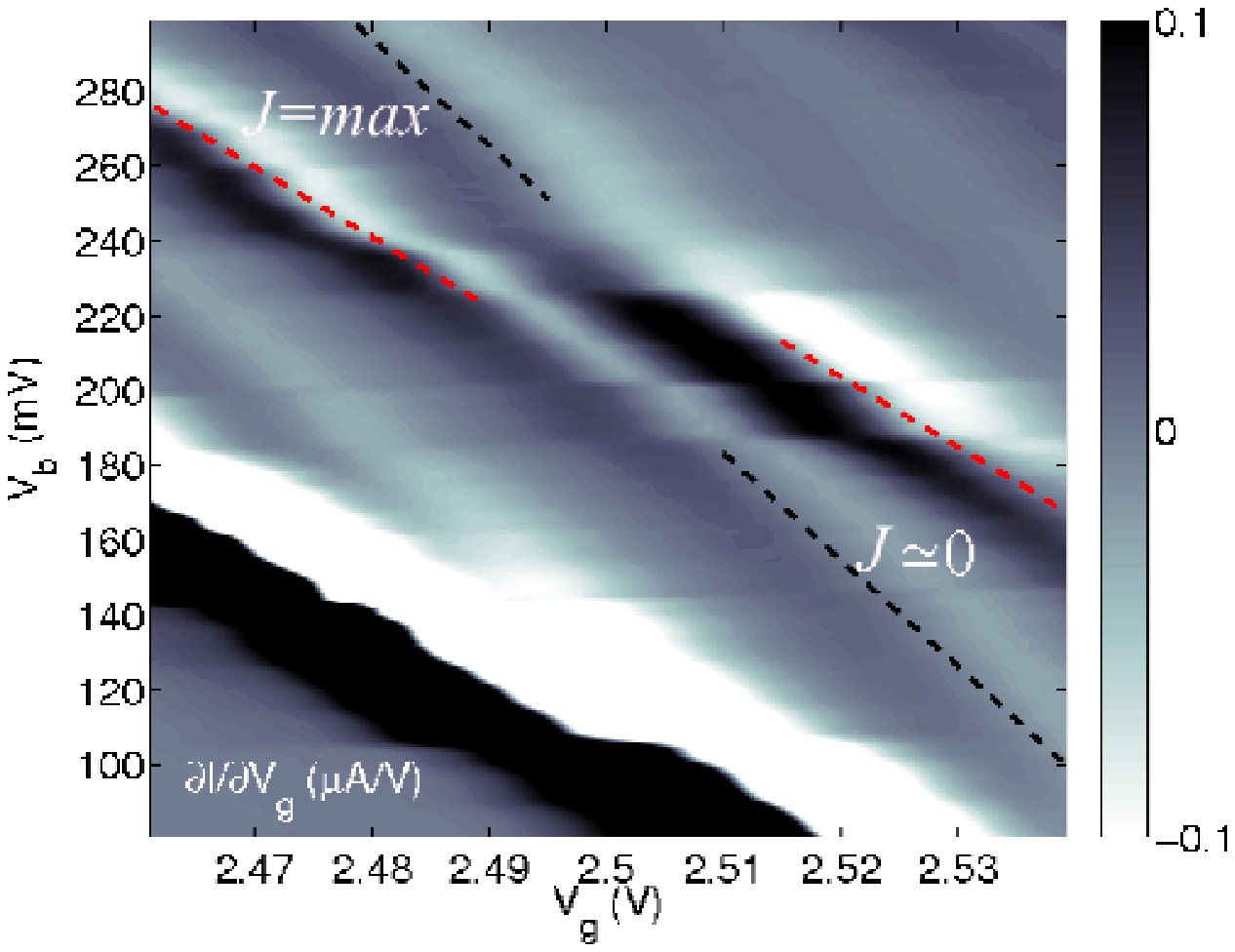}
\includegraphics[width=0.5\textwidth]{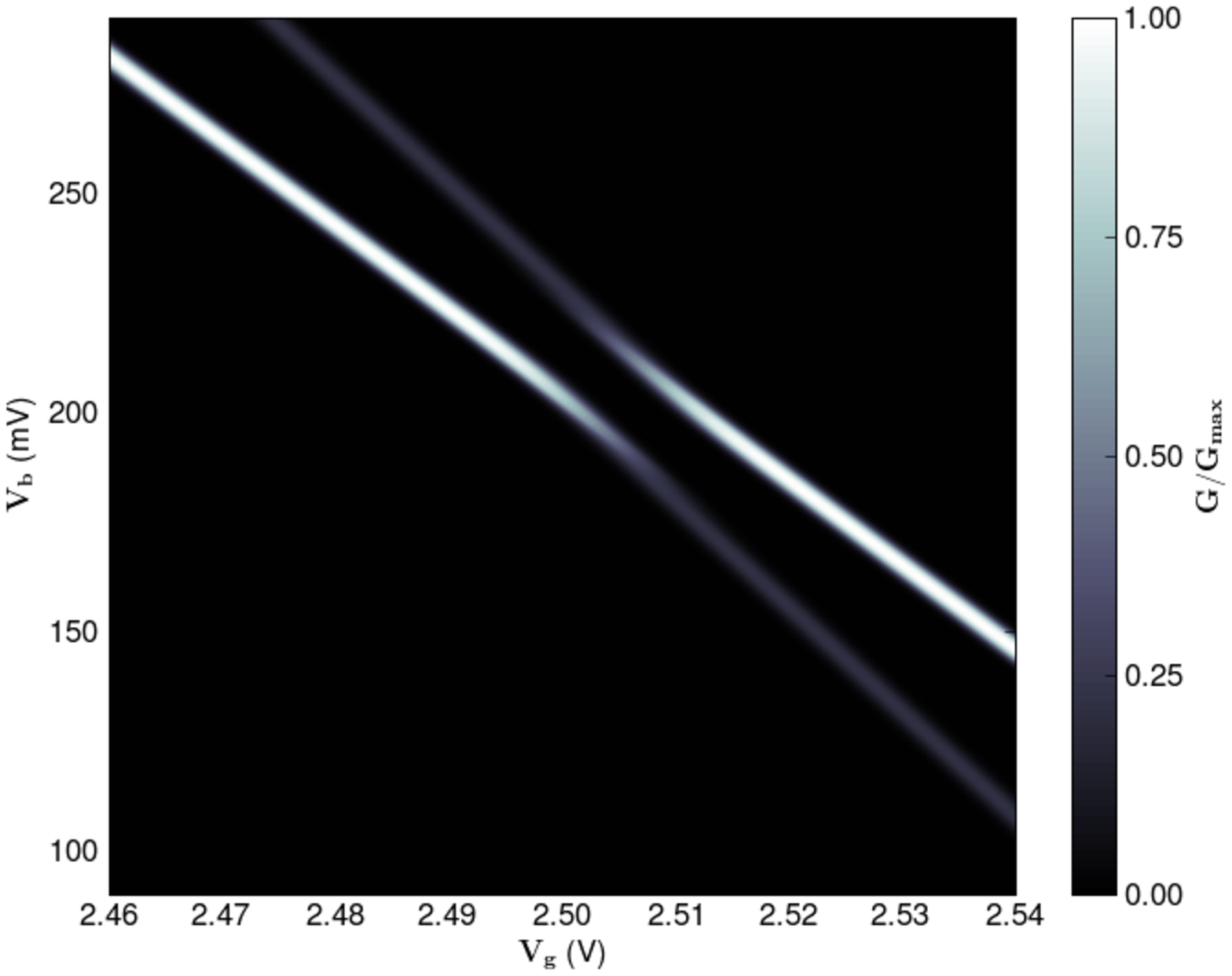}
\end{center}
\caption{\textbf{a}. The honeycomb region between the $D^{3-}$ and a quantum dot state. Contrary to the states at a lower energy of the As atom, a finite capacitance $C_{{As-QD}}$ among the quantum dot and the donor system determines the anticrossing of the two Coulomb blockade patterns at high gate volgate. The Coulomb blockade of the donor is guided with the red border, the quantum dot with the black. The adiabatic transition of the position of one electron from the quantum dot to the donor and \textit{viceversa} changes the exchange coupling $J$ of its spin with the three electrons in the donor from a negligible (right side) to a maximum (left side) value. 
 \textbf{b}. The simulation of the honeycomb region has been obtained by adding to the previous model a mutual capacitance $C_{{As-QD}}=5$ aF between the donor and the quantum dot.}
\label{Figure5}
\end{figure}

\end{document}